\documentclass[11pt]{article}
\usepackage{amsmath,amssymb,color,verbatim}
\usepackage{amsfonts}

\usepackage[pdftex]{graphicx} 

\makeatletter
\@addtoreset{equation}{section}
\makeatother

\newcommand{\be}{\begin{equation}}
\newcommand{\ee}{\end{equation}}
\newcommand{\bea}{\setlength\arraycolsep{2pt} \begin{eqnarray}}
\newcommand{\eea}{\end{eqnarray}}

\def\0{{\sst{(0)}}}
\def\1{{\sst{(1)}}}\def\2{{\sst{(2)}}}
 \def\3{{\sst{(3)}}}
\def\4{{\sst{(4)}}}
\def\5{{\sst{(5)}}}
\def\6{{\sst{(6)}}}
\def\7{{\sst{(7)}}}
\def\8{{\sst{(8)}}}
\def\sst#1{{\scriptscriptstyle #1}}

\def\ben{\begin{equation}}
\def\een{\end{equation}}
\def\half{{1 \over 2}}
\def\bea{\begin{eqnarray}}
\def\eea{\end{eqnarray}}

\DeclareMathOperator{\sech}{sech}

\begin{document}

\begin{flushright}
\hfill UPR-1266-T\ \ \ \ 
\end{flushright}

\vspace{25pt}
\begin{center}
{\Large {\bf Vacuum Polarization of STU Black Holes and their Subtracted Geometry Limit}}

\vspace{15pt}
{\Large Mirjam Cveti\v c$^{1,2}$, Gary W. Gibbons$^3$,  Zain H. Saleem$^{1,4}$,  Alejandro Satz$^1$}

\vspace{5pt}

{\it $^1$Department of Physics and Astronomy,\\
 University of Pennsylvania, Philadelphia, PA 19104, USA}

\vspace{5pt}
 
{\it $^2$Center for Applied Mathematics and Theoretical Physics,\\
University of Maribor, Maribor, Slovenia}

 \vspace{5pt}
{\it $^3$Department of Applied Mathematics and Theoretical Physics,\\
Wilberforce road, Cambridge, UK}

 \vspace{5pt}
{\it $^4$ National Center for Physics, Quaid-e-Azam University, Shahdara Valley Road, Islamabad, Pakistan}

\end{center}
\vspace{30pt}

\begin{abstract}
We study the vacuum polarization of a massless minimally coupled scalar field at the horizon of four-charge STU black holes. We compare the results for the standard asymptotically flat black holes and for the black holes obtained in the ``subtracted limit'', both in the general static case and at the horizon pole for the general rotating case. The original and the subtracted results are identical only in the BPS limit, and have opposite sign in the extremal Kerr limit. We also compute the vacuum polarization on the static solutions that interpolate between both the original and the subtracted case through a solution-generating transformation and show that the vacuum polarization stays positive throughout the interpolating solution. In the Appendix we provide a closed-form solution for the Green's function on general (static or rotating) subtracted black hole geometries.
\end{abstract}

\newpage
\tableofcontents

\section{Introduction}
Black Holes emit radiation and lose mass \cite{Hawking:1974sw}. This apparent violation of the Hawking area theorem can be explained by concluding that the energy flux out of the black hole is accompanied by an energy flux into the black hole across its horizon. This can happen because the polarization of the vacuum under the influence of the gravitational field can increase or decrease the local energy density of the zero-point fluctuations. The expectation value of the scalar field $\langle\varphi^2\rangle$, also called the vacuum polarization of the field, encodes much of the information on these quantum fluctuations. Computations of $\langle\varphi^2\rangle$ are a valuable tool in quantum field theory in curved spacetime, not only on their own regard as a measure of field fluctuations, but also as a tool for studying symmetry breaking effects and as a preliminary step in investigations of the stress-energy tensor and the Casimir effect.

 Candelas studied the effect of vacuum polarization on a scalar field in the Schwarzschild black hole background  \cite{Candelas:1980zt}  and  was able to calculate an analytical expression for $\langle\varphi^2\rangle $ at the horizon. Candelas' methods extend easily to charged static black holes, but the case of the rotating black hole is much more challenging; Frolov \cite{Frolov:1982pi} was able to calculate the analytical expression for $\langle\varphi^2\rangle $ only at the pole of the event horizon. There have also been numerical investigations of $\langle\varphi^2\rangle$ on general static black hole backgrounds beyond the event horizon (e.g. \cite{Anderson:1990jh} for asymptotically flat solutions and \cite{Flachi:2008sr} for the asymptotically anti-de Sitter case), and analytical computations at the horizon of a black hole threaded with a cosmic string \cite{Ottewill:2010hr}.

In this work we will focus on the vacuum polarization of the massless minimally coupled scalar field for  subtracted geometry black hole backgrounds. Subtracted geometry black holes are solutions of the bosonic sector of N=2 STU supergravity coupled to three vector multiplets. They are obtained when one subtracts certain terms from the warp factor of the general black holes \cite{CL11I,Cvetic:2011dn,Cvetic:1997xv}. This subtraction procedure when applied to the "original" N=2 STU black holes \cite{cvetic1} modifies their ``warp factor'' in such a way that wave equation becomes separable. The metric, however, remains a solution of the STU equations of motion. The subtracted metric can also be obtained from the original one through a scaling limit \cite{Cvetic:2012tr}, or through a continuously interpolating procedure using a solution-generating transformation \cite{Virmani:2012kw,Baggio:2012db,Cvetic:2013cja}. The subtracted geometry is asymptotically conical instead of being asymptotically flat. The subtracted black holes also display a Lifhshitz-like symmetry at the boundary and may be interpreted as being confined in an asymptotically conical box. 

 The modified black holes have the same horizon area and periodicity of the angular and time co-ordinates as the original black holes. The subtracted black hole geometry is a good approximation to the original one from the horizon to within the radius of the circular photon orbits of the original black holes. Quantum effects on curved spacetime, nevertheless, are often sensitive to non-local aspects of the geometry, and thus it is worth investigating whether the vacuum polarization at the horizon in the subtracted black holes is similar to the original values, or significantly different from them. Differences in quantum effects between original and subtracted black holes were already found in \cite{Cvetic:2014tka} for the horizon entanglement entropy at the subleading order.

This paper is organized as follows: In section 2 we will introduce our notation for describing both the original black holes and their subtracted counterparts. In section 3 we will calculate the analytical expressions for  $\langle\varphi^2\rangle$ for subtracted geometry black holes and compare it to the previously known value of the original black holes. In the static case we will calculate the vacuum expectation value at the horizon; for the rotating case we will calculate the value only at the pole of the horizon. Here we will also show that the values of $\langle\varphi^2\rangle$ in the original and subtracted geometries are related by the scaling limit mentioned above. We will also dedicate special discussion to the case of extremal black holes, showing that the original and subtracted results coincide for extremal static black holes (BPS limit) but not for extremal rotating black holes.
In section 4 we will calculate the vacuum expectation value for the interpolating  solution from the original black hole to the subtracted limit obtained through the solution-generating transformations. Section 5 contains the summary and discussion. In Appendix A we give the full expression for the Green's functions of the subtracted geometry black holes valid outside the horizon, which may be useful in later studies of, e.g., the vacuum stress-energy tensor and the self force problem. Im Appendix B we provide the vacuum polarization at the horizon pole for a rotating, non-subtracted STU black hole with four charges.

Throughout this paper we use units with $c=\hbar=G_N=1$.

\section{Black hole subtracted geometry}

 The general four-dimensional axisymmetric black hole metric is given by 
\begin{equation}
d  s^2  = -  \Delta^{-1/2}  G \, 
( d{ t}+{ {\cal  A\, \mathrm{d}\phi}})^2 + { \Delta}^{1/2}
\left(\frac{d r^2} { X} + 
d\theta^2 + \frac{ X}{  G} \sin^2\theta\, d\phi^2 \right)\,.\label{metric4d}
\end{equation}
Here the quantities $X, G, {\cal A},  \Delta $ are all functions of $r$ and $\sin \theta$ only (and depend on  the mass, rotation and charge parameters). The first three are the same for the original and the corresponding subtracted black hole,  while the difference in $\Delta$ is the hallmark of subtracted geometry. (The function $\Delta(r,\theta)$ is called the warp factor of the black hole geometry.) The physical parameters (mass $M$, angular momentum $J$ and charges $Q_I$) of the general four-charge black hole are parametrized in terms of auxiliary constants $m, a, \delta_I$ as:
\begin{eqnarray}
M  &= &{\frac{1}{4}}m\sum_{I=0}^3\cosh 2\delta_I ~, \label{Mdef} \nonumber\\
Q_I & = &{\frac{1}{4}}m \sinh 2\delta_I~,~(I=0,1,2,3)~,\nonumber\\
J & = &m\, a \,(\Pi_c - \Pi_s)~,
\end{eqnarray}
where  we employ the abbreviations
\begin{equation}
\Pi_c \equiv \prod_{I=0}^3\cosh\delta_I 
~,~~~ \Pi_s \equiv  \prod_{I=0}^3 \sinh\delta_I~.
\end{equation}
The functions  $X, G, {\cal A} $ are given by:
\begin{eqnarray}
{ X} & =& { r}^2 - 2{ m}{ r} + { a}^2~,\cr
{ G} & = &{ r}^2 - 2{ m}{ r} + { a}^2 \cos^2\theta\,, \cr
%&=& X-a^2\sin^2\theta\,,\cr
{ {\cal A}}  &=&{2{ m} { a} \sin^2\theta \over { G}}
\left[ ({ \Pi_c} - { \Pi_s}){  r} + 2{ m}{ \Pi}_s\right]\,.
  \label{c4d}
\end{eqnarray}
For the original black hole solutions, the remaining function $\Delta=\Delta_0$ is given by:
\begin{align}\label{delta0}
{ \Delta}_0 &= \prod_{I=0}^4 ({ r} + 2{ m}\sinh^2 { \delta}_I)
+ 2 { a}^2 \cos^2\theta \left[{ r}^2 + { m}{ r}\sum_{I=0}^3\sinh^2{ \delta_I}
+\,  4{ m}^2 ({ \Pi}_c - { \Pi}_s){ \Pi}_s \right. \nonumber\\ 
& \left.-  2{ m}^2 \sum_{I<J<K}
\sinh^2 { \delta}_I\sinh^2 { \delta}_J\sinh^2 { \delta}_K\right]
+ { a}^4 \cos^4\theta\,.
\end{align}
The particular case of $\delta_I=\delta$ for all $I$ recovers the usual Kerr-Newman black hole in a different parametrization. The matter sources for these black holes solutions are given in \cite{cvetic1,Chong:2005hr}.\footnote{The full solution with four electric charges and four magnetic charges is given in \cite{Chow:2014cca}. A subtracted version of this geometry was constructed in \cite{Cvetic:2014sxa}.}

The subtracted geometry is defined by the replacement of the function $\Delta_0$ by $\Delta_{sub}$ \cite{Cvetic:1997xv}, given by:

\begin{equation}\label{deltasub}
\Delta_{sub} = (2 m)^3 r ( \Pi_c^2 -\Pi_s^2) + (2m)^4 \Pi_s^2 - (2m)^2 ( \Pi_c-\Pi_s)^2 a^2 \cos^2 \theta\,.
\end{equation}
The different scaling of $\Delta$ at $r\to\infty$ (namely, the dominant terms being $\sim r$ instead of $\sim r^4$) is what makes the subtracted geometry asymptotically Lifhshitz-like instead of asymptotically flat.

The subtracted metric is a solution of the bosonic sector four-dimensional ${\cal N}=2$ supergravity coupled to three vector supermultiplets. We will not require for our current purposes the detailed form of matter fields supporting the geometry, which are given in \cite{Cvetic:2011dn}. In \cite{Cvetic:2012tr} it was shown that one can obtain the subtracted geometry through a scaling limit, making in the original black hole metric the redefinitions
\begin{align}\label{scalinglimit}
\,&r\to r\varepsilon\,,\quad\quad t\to t\varepsilon^{-1}\,,\quad\quad m\to m \varepsilon\,,\quad\quad a\to a \varepsilon\,,\nonumber\\
 &\sinh^2 \delta_0\to \frac{\Pi_s^2}{ \Pi_c^2-\Pi_s^2}\,,\quad\quad\sinh^2 \delta_ I\to \varepsilon^{-4/3} (\Pi_c^2-\Pi_s^2)^{1/3}\,,
\end{align}
and taking the $\varepsilon\to 0$ limit.

In both the original and the subtracted case the horizons, specified by $X=0$, are at:
\begin{equation}\label{defhorizons}
r_\pm=m\pm \sqrt{m^2-a^2}\, .
\end{equation}
The inverse surface gravity at each horizon is given by:
\begin{equation}\label{defkappa}
\frac{1}{\kappa_\pm}=2m\left[\frac{m}{\sqrt{m^2-a^2}}(\Pi_c+\Pi_s)\pm(\Pi_c-\Pi_s)\right]\,.
\end{equation}
The temperature in the Hartle-Hawking state is given by $T=(2\pi)^{-1}\kappa$, where $\kappa\equiv\kappa_+$.
We also define the angular velocities:
\begin{equation}
\Omega_\pm=\kappa_\pm\frac{a}{\sqrt{m^2-a^2}}\,.
\end{equation}
The subtracted versions of the Kerr-Newman, Kerr, Reissner-Nordstr\"om, and Schwarzschild black holes are obtained, respectively, by setting all $\delta_I=\delta$, setting all $\delta_I=0$, setting all $\delta_I=\delta$ and $a=0$ , and setting all $\delta_I=a=0$. Though the horizon surface gravity and area of each subtracted black hole match the corresponding original (=non-subtracted) ones, other properties of the geometry are different. In particular, due to the nontrivial supporting matter fields, the Ricci tensor does not vanish for any of the black holes under consideration, and the Ricci scalar vanishes at the horizon, but not at an arbitrary point.

\section{Vacuum polarization at the black hole event horizon}

 \subsection{Green's function and counterterms}
 
There are no existing analytic tools for computing the vacuum polarization on the entire horizon of rotating black holes. Therefore we will restrict our attention to the general static case, and to $\langle \varphi^2 \rangle$ at the pole ($\theta=0$) of the horizon of rotating black holes. The vacuum polarization for the original Schwarzschild and (at the pole) Kerr-Newman black holes have been computed in  \cite{Candelas:1980zt,Frolov:1982pi} . We will add to these pre-existing computations the original four-charge black hole, and compare all the results with the corresponding subtracted cases. All our calculations will assume a massless, minimally coupled scalar field.

The algorithm for computing the vacuum polarization $\langle \varphi^2 \rangle$ of a scalar field  in a  thermal state on a curved background is simple in principle, and is composed of two steps. The first one is to compute the Euclidean Green's function $G(x,x')$, by solving the wave equation on the Wick-rotated Euclidean manifold with the time periodicity corresponding to the temperature of the field. The second one is to take the coincidence limit of the Green function, regularizing the UV divergences by subtracting appropriate counterterms which depend on the geodesic distance $s(x,x')$. In our case we will always implement radial separation, setting $t=t'$, $\varphi=\varphi'$, $\theta=\theta'$ ($=0$ in the rotating case) and $r=r_+, r' = r_++\epsilon$.

The  wave equation for the Euclidean Green's function, after Wick rotating $t\to-i\tau$, is:
\begin{equation}\label{greeneq1}
\Box \; G_H( -i\tau , x , \theta , \phi \; ; {-i \tau}' , {x}' , {\theta}' , {\phi}') = -i \frac{1}{(r_+-r_-)}\frac{1 }{ \sqrt{-g}} \delta ( \tau - {\tau}') \delta ( x-x') \delta ( \Omega , \Omega')\,,
\end{equation}
where $x =(r - \frac{1}{2}(r_+ + r_-))/( r_+ - r_-)$ is a convenient rescaled radial variable, and $g$ is the determinant of the original metric (\ref{metric4d}). For subtracted geometry, the solution in the general rotating case is expanded as:
\begin{align}\label{greenexprot}
G_H( -i\tau , x , \theta , \phi \; ; {-i \tau}' , {x}' , {\theta}' , {\phi}')&= \frac{1}{r_+-r_-} \frac{ i \kappa}{ 2 \pi} \sum_{n =- \infty}^ {\infty} e^{ i n \kappa( \tau -{\tau}')} \sum_{l=0}^{\infty}\frac{(2l+1)}{4\pi} \nonumber\\
&\times\sum_{m =- l}^ {l} \frac{(l-m)!}{(l+m)!}e^{ i m ( \phi -{\phi}')} P_l^m( \cos{\theta})P_l^m( \cos{\theta'}) G _{mln}(x,x')\,.
\end{align} 
In the original case the expansion involves Lam\'e functions instead of the Legendre functions $P_l^m$.
In the general static case we may, by symmetry, omit the sum over the magnetic quantum number $m$ (not to be confused with the black hole mass parameter) and write directly:
\begin{equation}\label{expansionst}
G_H( -i\tau , x , \theta , \phi \; ; {-i \tau}' , {x}' , {\theta}' , {\phi}')= \frac{1}{2m} \frac{i\kappa}{ 2 \pi} \sum_{n =- \infty}^ {\infty} e^{ i n \kappa( \tau -{\tau}')}\sum_l \frac{ (2 l+1)}{4\pi} P_l ( \cos \Theta) G _{ln}(x,x')\,,
\end{equation} 
where $\Theta$ is the angle between the two points. Expansion (\ref{expansionst}) is valid in the general static case (both original and subtracted).

The radial wave equation that $G _{mln}(x,x')$ or $G_{ln}(x,x')$ satisfy is not solvable in closed form for the original black holes. For subtracted black holes there exist closed-form solutions in terms of hypergeometric functions, which we provide in the Appendix. It turns out that in both the subtracted and original cases, the radial Green's function with one point at the horizon vanishes except when $n=0$ (static case) or $n=m=0$ (rotating case). Therefore we only need $G_{ln}(x,x')$ and $G _{0l0}(x,x')$, which are the same in all cases:
\begin{equation}\label{green0gen}
G_{l0}(x,x')=G_{0l0}(x,x')=2\left[P_l(2x)Q_l(2x')H(x'-x)+P_l(2x')Q_l(2x)H(x-x')\right]\,,
\end{equation}
Here $P_l,Q_l$ are the Legendre polynomials and the Legendre functions of the second kinds respectively, and $H$ is the Heaviside step function. The external horizon $r=r_+$ is at $x=1/2$. It follows that in the general static case we have:
\begin{equation}
G(-i\tau , r_+, \theta,\phi;-i\tau ', r_++\epsilon, \theta,\phi ) = \frac{i\kappa}{2\pi}\frac{1}{4\pi m}\sum_{l=0}^{\infty} (2l+1)Q_l\left(\frac{m+\epsilon}{m}\right)\,,
\end{equation}
and in the rotating case we have:
\begin{equation}
G_H( -i\tau , r_+ , 0, \phi \; ; {-i \tau}' , r_+ + \epsilon ,0, {\phi})= \frac{ i \kappa}{ 2 \pi} \frac{1}{2\pi r_0} \sum_{l=0}^{\infty}(2l+1)Q_l\left(\frac{2\epsilon+r_0}{r_0}\right)\,,
\end{equation} 
where $r_0=r_+-r_-=2\sqrt{m^2-a^2}$; the latter expression clearly reduces to the former one for $a=0$. The sums are evaluated using the Heine Identity \cite{whittaker}:
\begin{equation}
\sum_{l=0}^{\infty} (2 l+1) P_{l}(\Psi)Q_{l}(\zeta)=\frac{1}{(\zeta-\Psi)}\,.
\end{equation}
 Therefore in the general static case we have:
\begin{equation}\label{horizongreen}
G(-i\tau , r_+, \theta,\phi;-i\tau , r_++\epsilon, \theta,\phi ) = \frac{i\kappa}{8\pi^2\,\epsilon}\,,
\end{equation}
and the same expression holds for $\theta=0$ in the general rotating case. Thus the results of Candelas and Frolov \cite{Candelas:1980zt,Frolov:1982pi} for the Green function on Schwarzschild and Kerr-Newman generalize to black holes with four charges and with or without subtracted geometry.

The vacuum polarization is given by the limit:
\begin{align}\label{vacpollimit}
\langle\varphi^2\rangle_{r_+}&= \lim_{\epsilon\to0^+} \left[-i G(t, r_+, \theta,\phi;t, r_++\epsilon, \theta,\phi ) -\frac{1}{8\pi^2 \sigma}-\frac{1}{96\pi^2}\frac{R_{ab}\,\sigma^{,a}\sigma^{,b}}{\sigma}\right]\nonumber\\
&= \lim_{\epsilon\to0^+} \left[ \frac{\kappa}{8\pi^2\,\epsilon} -\frac{1}{8\pi^2 \sigma}-\frac{1}{96\pi^2}\frac{R_{ab}\,\sigma^{,a}\sigma^{,b}}{\sigma}\right]\,,
\end{align}
where $\sigma = \frac{1}{2}s^2(r_+,r_++\epsilon)$ is half of the geodesic distance squared between the points $(t, r_+, \theta,\phi)$ and $(t, r_++\epsilon, \theta,\phi )$ (with $\theta=0$ in the rotating case). The two counterterms are the non-vanishing parts of the Hadamard expansion of the Green's function \cite{Christensen:1976vb}.

It is seen that even though the Green's function term for subtracted black holes matches the non-subtracted one, the vacuum polarizations in both cases will not coincide. This is because the counterterms that need to be  subtracted from $-i G$ have nontrivial dependence on the warp factor at the subleading order that survives the cancelation of divergences. For example, the second counterterm vanishes in vacuum solutions like the original Schwarzschild and Kerr black holes, yet it does not vanish for any of the subtracted black holes.

The sum of the two counterterms in (\ref{vacpollimit}) must be computed up to order $O(1)$ in an expansion in powers of $\epsilon$. To achieve this, we write the geodesic distance as:

\begin{equation}
s(r_+,r_++\epsilon)=\int_{r_+}^{r_++\epsilon}\mathrm{d}r\, (g_{rr})^{1/2} = \int_{r_+}^{r_++\epsilon}\mathrm{d}r\,\frac{\Delta^{1/4}(r)}{X^{1/2}(r)}\,.
\end{equation}
In the rotating case, the integral is computed by evaluating the metric function $\Delta$ at arbitrary $\theta$ setting $\theta=0$ at the end of the calculation. Even though the integral cannot in general be computed in closed form, it is possible to expand the integrand in order to compute $\sigma$ accurately to the required order in $\epsilon$. 

Once we have obtained  $\sigma(\epsilon)$, we can easily compute the second counterterm $R_{ab}\sigma^{,a}\sigma^{,b}{\sigma}^{-1}$ by using the fact that for radial separation, both in the static case and at the pole in the rotating case, $\sigma^{,r}$ is the only nonvanishing component of $\sigma^{,a}$ and we have
\begin{equation}
\sigma = \frac{g_{rr}}{2}\left(\sigma^{,r}\right)^2\,.
\end{equation}

The preceding collection of formulas make straightforward the computation of $\langle\varphi^2\rangle_{r_+}$ in the general case. The result may be succinctly expressed as:
\begin{equation}\label{gaussian}
\langle\varphi^2\rangle_{r_+}=\frac{1}{48\pi^2}\left(K_0+R^r_{\,\,r}\right)\,.
\end{equation}
Here $K_0$  is the intrinsic curvature of the horizon (its Gaussian curvature as a two-dimensional surface, evaluated at the pole in the rotating case). The first term of (\ref{gaussian}) was derived originally by Frolov \cite{Frolov:1985yu} based on an earlier approximation scheme by Page \cite{Page:1982fm}. A covariant form of this expression is:
\begin{equation}\label{gaussiancov}
\langle\varphi^2\rangle_{r_+}=\frac{1}{48\pi^2}\left(K_0+\frac{1}{2}\left[\frac{R_{ab}\,\sigma^{,a}\sigma^{,b}}{\sigma}\right]\right)\,,
\end{equation}
where the square brackets denote the coincidence limit.\footnote{Note that the $R_{ab}$ term appears with the opposite sign here than in (\ref{vacpollimit}); this is because the finite part of the $\sigma^{-1}$ counterterm provides not only $K_0$ but also an additional contribution proportional to $R^r_{\,r}$.}

We give below the results first for each of the non-subtracted black holes, and then for their subtracted counterparts.

\subsection{Results for original black holes}

The vacuum polarization at the horizon for the general static four-charge original black hole is:

\begin{equation}\label{4qorig}
\langle\varphi^2\rangle_{r_+}^{4Q_{orig}}= \frac{\sum_I \sech^2{\delta_I}}{ 768 m^2 \pi^2 \Pi_c}\,.
\end{equation}

This result is novel, and reduces when $\delta_I=\delta$ and when $\delta_I=0$ to the previously known results for the vacuum polarization on the Reissner-Nordstr\"om and Schwarzschild black holes \cite{Candelas:1980zt,Frolov:1982pi}:
\begin{eqnarray}
\langle\varphi^2\rangle_{r_+}^{RN_{orig}}&=&\frac{1}{ 192 m^2 \pi^2\,\cosh^6\delta}\,, \\
\langle\varphi^2\rangle_{r_+}^{Sch_{orig}}&=& \frac{1}{ 192 m^2 \pi^2}\,.
\end{eqnarray}

The result at the pole of the general four-charge rotating black hole is readily obtainable as well. Because of its great length, we have included it in Appendix B. In the limit $\delta_I=\delta$, it reduces  to the Kerr-Newman result first derived in \cite{Frolov:1982pi}:
\begin{equation}\label{frolov}
\langle\varphi^2\rangle_{r_+,\,\theta=0}^{KN_{orig}}= \frac{m^2-2a^2+m\sqrt{m^2-a^2}\cosh 2\delta}{6m^2\pi^2\left(4\sqrt{m^2-a^2}\cosh 2\delta+m(3+\cosh 4\delta)\right)^2}\,.
\end{equation}
To verify the equality of this result with that of \cite{Frolov:1982pi} one must bear in mind that in our notation the standard mass and charge parameters become $M = m \cosh(2\delta)$ and $Q =  m \sinh(2\delta)$.

\subsection{Results for subtracted black holes}

The vacuum polarization at the horizon for the general static four-charge subtracted black hole is:
\begin{equation}\label{result4qsub}
\langle\varphi^2\rangle_{r_+}^{4Q_{sub}}=\frac{\Pi_c^2-\Pi_s^2}{768\pi^2 m^2\Pi_c^3}\,.
\end{equation}
In the particular cases of subtracted Reissner-Nordstr\"om and subtracted Schwarzschild we obtain:
\begin{eqnarray}
 \langle\varphi^2\rangle_{r_+}^{RN_{sub}}&=&\frac{1-\tanh^8{\delta}}{768\pi^2 m^2\cosh^4 \delta}\,,\\  \langle\varphi^2\rangle_{r_+}^{Sch_{sub}}&=&\frac{1}{768\pi^2 m^2}\,.
\end{eqnarray}

Comparing with the corresponding results for non-subtracted black holes, we see that all the results for the horizon vacuum polarization on the static subtracted geometries differ from their original counterparts. However, the difference is quantitative and not qualitative. The sign of the result, and the general way it behaves as a function of the parameters, is generally unchanged.

The vacuum polarization at the pole of the fully general four-charge rotating subtracted black hole is relatively simpler than its corresponding non-subtracted expression given in Appendix B. It reads:
\begin{align}
\langle\varphi^2\rangle_{r_+,\,\theta=0}^{gen_{sub}}&=\frac{\Pi_c-\Pi_s}{192m\pi^2\left[m\left(2m(\Pi_c^2+\Pi_s^2)+r_0(\Pi_c^2-\Pi_s^2)\right)-a^2(\Pi_c-\Pi_s)^2\right]^{5/2}}\times\nonumber\\
&\Big[m\, a^2(\Pi_c-\Pi_s)\left(r_0(\Pi_c^2-\Pi_s^2)-8m\,\Pi_c\,\Pi_s\right)\nonumber\\
&+2m^3(\Pi_c+\Pi_s)\left(2m(\Pi_c^2-\Pi_s^2)+r_0(\Pi_c^2+\Pi_s^2)\right)-2a^4(\Pi_c-\Pi_s)^3\Big]\,,
\end{align}
where as before we use the notation $r_0=2\sqrt{m^2-a^2}$. In the Kerr-Newman limit $\delta_I=\delta$ we obtain:
\begin{align}
\,&\langle\varphi^2\rangle_{r_+,\,\theta=0}^{KN_{sub}}=\nonumber\\
&\frac{\left(c^4-s^4\right)^4\left(a^2m(c^8r_0-8c^4s^4m-s^8r_0)-2a^4\right)+2m^3(c^8-s^8)\left(c^8(r_0+2m)+s^8(ro-2m)\right)}{192m\pi^2\left[m(r_0+2m)c^8 +m(2m-r_0)s^8-a^2(c^4-s^4)^2\right]^{5/2}}\,.
\end{align}
where for compactness we write $c,s$ for $\cosh\delta$ and $\sinh\delta$. These results reduce to the above formulas for static black holes when $a=0$. In this case, the subtracted expression turns out to be significantly less simple than Frolov's original result (\ref{frolov}). It also presents the qualitative difference of being positive for all values of the parameters, whereas the original result can be vanishing or negative.

To conclude this section, we remark that there exists an alternative derivation of the vacuum polarization for subtracted black holes. Instead of redoing the calculation  in (\ref{vacpollimit}) for the new geometries, we can apply the scaling limit (\ref{rescaling}) directly to the original black hole results of Section 3.1. We have confirmed  that performing this transformation on $\langle\varphi^2\rangle_{r_+}^{orig}$ in fact results in the expressions we gave in this section for $\langle\varphi^2\rangle_{r_+}^{sub}$. 

\subsection{Extremal black holes}

There is a limit in which the subtracted geometry and the original geometry coincide. This is the so-called BPS limit, which consists in rescaling the parameters $m,a,\delta_I$ as follows:
\begin{eqnarray}
m&\to&m\varepsilon\,,\\
a&\to&a\varepsilon\,,\\
\mathrm{e}^{2\delta_I}&\to&\frac{1}{\varepsilon}\mathrm{e}^{2\delta_I}\,,
\end{eqnarray}
and taking the limit $\varepsilon\to 0$. It is easily seen that this limit results in $J=0$, $M=\sum_I Q_I$, $r_+=r_-$, and $\kappa=0$. Therefore the BPS limit describes an extremal static four-charge black hole. 

When taking this limit directly on both our original and our subtracted results, we obtain, as expected, the same limiting value:
\begin{equation}
\langle\varphi^2\rangle_{r_+}^{BPS}\to 0\,.
\end{equation}

\begin{figure}[h,t]
\begin{center}
\includegraphics[width=0.8\textwidth]{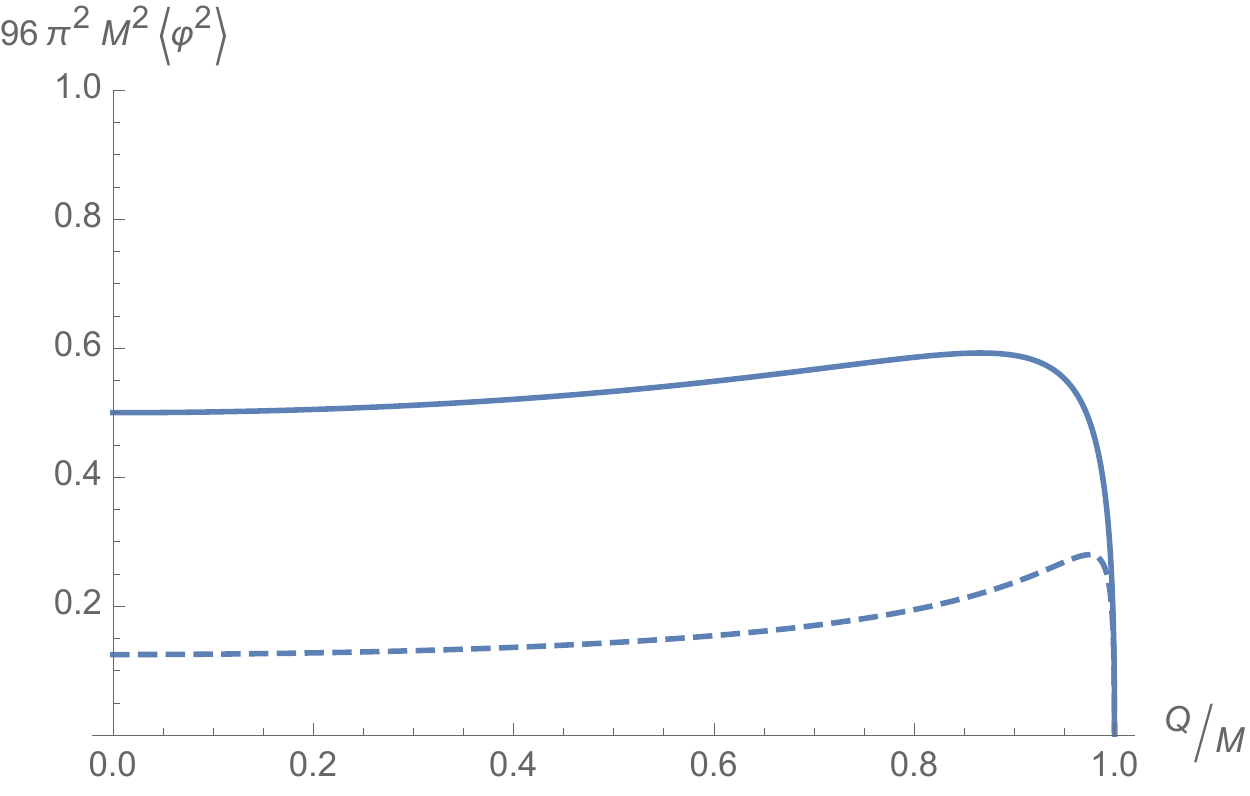}
\end{center}
\caption{Vacuum polarization at the horizon original Reissner-Nordstr\"om black hole (full line) and subtracted Reissner-Nordstr\"omr (dashed line). In the extremal limit both results coincide and vanish.}
\label{RNplot}
\end{figure}

The way the zero result in the extremal limit is achieved in the original black hole and the subtracted one, for the Reissner-Nordstr\"om case of all $\delta_I=\delta$, is plotted in Figure \ref{RNplot} as a function of the ratio of physical charge ($Q=m\sinh\delta$) to physical mass ($M=m\cosh\delta$).

We remark in passing that the BPS limit does not make each term of the expression (\ref{gaussian}) vanish separately; rather, in the BPS limit both terms are finite and of opposite value:
\begin{equation}
K_0=\frac{1}{4M^2}=-R^r_{\,r\,}\,\quad\quad\quad \mathrm{(BPS\,\,limit)}\,.
\end{equation}

It is tempting to interpret the vanishing of the vacuum polarization in the BPS limit as related to the zero temperature of the extremal black hole and as evidence of the intrinsically thermal nature of the field fluctuations. However, this interpretation fails to account for the fact that for extremal rotating black holes the vacuum polarization does not vanish. Setting $m=a$ results also in an extremal black hole with zero temperature, as can be seen from (\ref{defhorizons}) and (\ref{defkappa}), both for original and subtracted black holes and for any value of  $\delta_I$. Looking for simplicity just at the pure Kerr case $\delta_I=0$, the original vacuum polarization at the pole in this limit is given by
\begin{equation}
\langle\varphi^2\rangle_{r_+,\,\theta=0}^{Kerr_{ext}}=-\frac{1}{96\pi^2m^2}\,,
\end{equation}
and the corresponding subtracted value is given by
\begin{equation}
\langle\varphi^2\rangle_{r_+\,\theta=0}^{Kerr_{ext-sub}}=\frac{1}{96\pi^2m^2}\,,
\end{equation}
that is, exactly opposite in sign. We do not have an explanation for this fact, which could be a mere numerical coincidence. 

\begin{figure}[h,t]
\begin{center}
\includegraphics[width=0.8\textwidth]{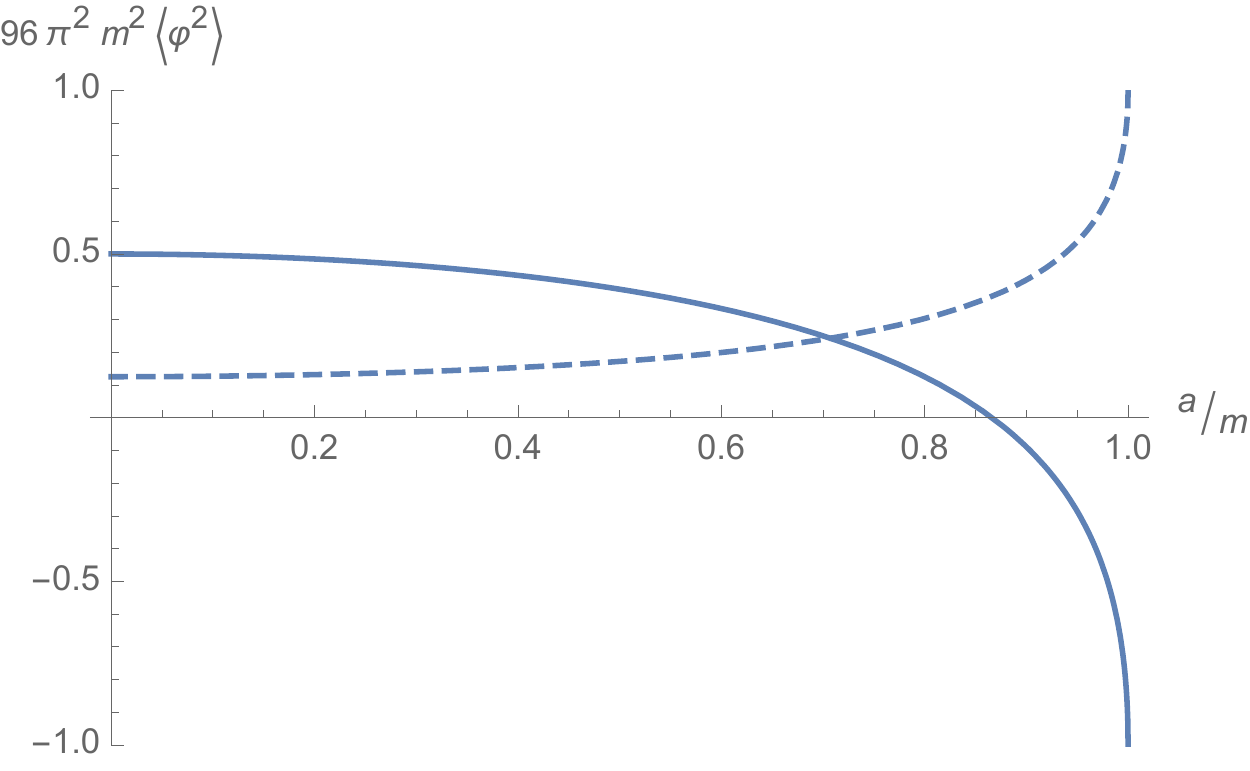}
\end{center}
\caption{Vacuum polarization at the horizon pole for the original Kerr (full line) and subtracted Kerr (dashed line). In the extremal limit the results are equal in magnitude and opposite in sign.}
\label{kerrplot}
\end{figure}

The values of $\langle\varphi^2\rangle_{r_+,\,\theta=0}$ for both kinds of Kerr black holes are not so simply related in the non-extremal case; the original one exhibits a zero value for $a/m=\sqrt{3}/2$ (as first noted by Frolov \cite{Frolov:1982pi}) while the subtracted one never vanishes. As a function of the dimensionless variable $y=a/m$, these results read:
\begin{equation}
\langle\varphi^2\rangle_{r_+,\,\theta=0}^{Kerr_{orig}}=-\frac{1}{96\pi^2m^2}\frac{3-2y^2-3\sqrt{1-y^2}}{y^2}\,,
\end{equation}
\begin{equation}
\langle\varphi^2\rangle_{r_+\,\theta=0}^{Kerr_{sub}}=\frac{1}{96\pi^2m^2}\frac{2-y^2+(2+y^2)\sqrt{1-y^2}}{\left[2-y^2+2\sqrt{1-y^2}\right]^{5/2}}\,.
\end{equation}
These results are contrasted in Figure \ref{kerrplot}. The change in sign of $\langle\varphi^2\rangle_{r_+,\,\theta=0}^{Kerr_{orig}}$ tracks directly the change of sign of the intrinsic curvature $K_0$ of the near-extremal Kerr event horizon that was first indicated by Smarr \cite{Smarr:1973zz}. This is because for the original Kerr black hole the second term of (\ref{gaussian}) vanishes. In the subtracted geometry case, both terms of (\ref{gaussian}) contribute and the result is always positive.

\section{Vacuum polarization for interpolating static black holes}

In this section we will use an interpolating member of the four-charge family of static black hole solutions to find a formula for the vacuum polarization that interpolates between the original black hole value, its subtracted geometry value, and its value in the BPS limit (which, as we have seen, is zero).

Stationary solutions of the STU supergravity theory are acted on by the group $O(4,4)$. In particular, the   original (asymptotically flat) black holes may all be obtained  by acting with a $O(1,1)^4$ subset of solution-generating transformations acting on a neutral stationary black hole. These transformations are parametrized by four boosts $\delta_I$ and one may obtain asymptotically flat BPS solutions in the limit $\delta_I\to\infty$. %All solutions obtained through these transformations are asymptotically flat.

 In  \cite{Cvetic:2013cja}, it was shown that acting on any of the original black holes with a different $O(1,1)^4\subset O(4,4)$, parametrized by four $\alpha_I$ with $0<\alpha_I<1$, one obtains again all the asymptotically flat original black holes. However, if any any of the $\alpha_I$ parameters equals unity, we obtain further solutions which are not asymptotically flat. In particular, the subtracted geometries studied in these paper are obtained by setting all but one $\alpha_I$ equal to unity. If all $\alpha_I$ are set to unity, one obtains a Robertson-Bertotti-type solution, which coincide with the near-horizon geometries of the asymptotically flat BPS black holes. 

%This interpolating solution was constructed in \cite{Cvetic:2013cja}  using Harrison transformations. These are a family of transformations  that act on black hole solutions of the theory under consideration (the bosonic sector of N=2 STU supergravity coupled to three vector multiplets). They are  characterized by eight real parameters, but we need only a four-parameter subfamily for our present purposes. The values of the parameters, which we label $(\alpha_0, \alpha_i)$ with $i=1,2,3$, range between 0 and 1. The original black holes correspond to case where all the Harrison parameters are set to zero and the BPS limit corresponds to setting all the parameters to unity.\footnote{When all the four $\alpha_I$ parameters are set to unity, the four-dimensional static geometry becomes $\mathrm{AdS_2}\times S_2$. For details see \cite{Cvetic:2013cja}.} The subtracted limit requires setting $\alpha_i=1$ with the value of the fourth parameter $\alpha_0$ depending on the charges of the black hole.  

We shall restrict ourselves in this paper to $\alpha$-transformations acting within the subfamily of static black holes, because in this case the transformation linking an original black hole to its subtracted counterpart preserves the mass and charge parameter values along its orbit,

The $\alpha_I$ transformation only affects the warp factor $\Delta$ in the black hole metric. The warp factor for a static black hole with mass $m$ and charges $\delta_I$, as a function of the $\alpha_I$ parameters, is given by
\begin{eqnarray}
\Delta_{int}&=& \prod_{J=0}^4\left[(1-\alpha_J^2)r+2m(\alpha_J \cosh \delta_J + \sinh \delta_J)^2\right]\,\\
&\equiv& \prod_J\left[ a_J r + 2m\, b_J\right]\,,
\end{eqnarray}
where we define $a_J=1-\alpha_J^2$, $b_J=(\alpha_J \cosh \delta_J + \sinh \delta_J)^2$.

The static black hole subtracted geometry is obtained by setting:
\begin{equation}\label{subharrison}
\alpha_i = 1\,,\quad\quad\alpha_0=\alpha_0^*\equiv\frac{\Pi_s\cosh\delta_0-\Pi_c\sinh\delta_0}{\Pi_c\cosh\delta_0-\Pi_s\sinh\delta_0}\,,
\end{equation}
and doing in addition a rescaling of the metric given by:
\begin{equation}\label{rescaling}
g_{\mu\nu}\to \mathrm{e}^{-2c_0}g_{\mu\nu}\,,\quad\quad \mathrm{e}^{-2 c_0}=\frac{\Pi_c\cosh\delta_0-\Pi_s\sinh\delta_0}{\mathrm{e}^{\delta_1+\delta_2+\delta_3}}\,.
\end{equation}

The vacuum polarization of the interpolating four charge black holes is computed straightforwardly as before from (\ref{vacpollimit}) by computing the geodesic distance, its derivative and the Ricci tensor of the interpolating metrics. The result is written in general in the form:

\begin{eqnarray}
\langle\varphi^2\rangle_{r_+}^{4Q_{int}} =  \frac{4\;a_1a_2a_3a_4 + 3( a_1a_2a_3b_4+ ...) + 2 ( a_1a_2 b_3b_4+ ...) + 1( a_1 b_2 b_3 b_4+ ...)}{768 \pi^2 m^2\,\prod_J (a_J  +  b_J)^{3/2}}\,,
\end{eqnarray}
where the dots indicate all the inequivalent terms obtained by permuting indices. It is easily checked that the above expression reduces to the original result in the correct limit,
\begin{equation}
\langle\varphi^2\rangle_{r_+}^{4Q_{int}} = \langle\varphi^2\rangle_{r_+}^{4Q_{orig}}\quad\quad\quad\quad\quad \;\;(\alpha_J=0)\,,
\end{equation}
where the original result was given in (\ref{4qorig}). Since each $a_J$ vanishes when $\alpha_J=1$, the BPS limit correctly obtains the vanishing result we discussed above: 
\begin{equation}
\langle\varphi^2\rangle_{r_+}^{4Q_{int}} =\langle\varphi^2\rangle_{r_+}^{BPS}=0\;\;\;\;\;\;\;\;\;\;\;\;\;\; (\alpha_J=1).
\end{equation}
The value of $\langle\varphi^2\rangle_{r_+}^{4Q}$ in the subtracted geometry can be found by setting the $\alpha$-parameters to the values (\ref{subharrison}), and then performing a rescaling corresponding to (\ref{rescaling}) on the result, giving:
\begin{equation}
\langle\varphi^2\rangle_{r_+}^{4Q_{sub}}= \left.\langle\varphi^2\rangle_{r_+}^{4Q_{int}}\right|_{\alpha_i=1,\,\alpha_0=\alpha_0^*} \times e^{2c_0}\,.
\end{equation}
It is verified that this expression agrees with (\ref{result4qsub}).

The results become particularly simple for the Schwarzschild interpolating black hole:
\begin{equation}
\langle\varphi^2\rangle_{r_+}^{Sch_{int}} =  \frac{4- \alpha_0^2- \alpha_1^2 -\alpha_2^2 - \alpha_3^2 }{768 \pi^2 m^2}\,.
\end{equation}
In this case the subtracted limit has $\alpha_0^*=0$, and also $c_0=0$ (no rescaling). Thus we verify:
\begin{eqnarray}
\langle\varphi^2\rangle_{r_+}^{Sch_{int}} &=& \langle\varphi^2\rangle_{r_+}^{Sch_{orig}}\quad\quad\quad\,( \alpha_J=0)\,,\\&=& \langle\varphi^2\rangle_{r_+}^{Sch_{sub}}\quad\quad\quad\,\, ( \alpha_i=1, \; \alpha_0=0 )\,,\\
&=&0 \quad\quad\quad\quad\quad\quad\quad(\alpha_J=1)\,.
\end{eqnarray}

\section{Conclusions}

In this paper we have investigated the vacuum polarization $\langle\varphi^2\rangle$ for a wide class of black holes that are solutions of the  bosonic sector of N=2 STU supergravity coupled to three vector multiplets. These black holes are characterized by a mass parameter $m$, a rotation parameter $a$, and four charge parameters $\delta_I$. We have focused our attention on two general subclasses: the original black holes, which are asymptotically flat (and include the usual black holes of the Kerr-Newman family, in the limit where all charges coincide), and the subtracted black holes, which modify the warp factor of the metric (changing it from (\ref{delta0}) to (\ref{deltasub})) and are asymptotically Lifhshitz. The subtracted geometry is of special interest because it makes the wave equation separable in addition to providing a good approximation to the original geometry in the near-horizon regime.

We have computed the vacuum polarization of a massless, minimally coupled scalar field at the horizon (for static black holes) and at the pole of the horizon (for rotating black holes). The calculation was outlined in Section 3.1, using results for the Green's function that are derived in Appendix A. The results for original black holes are presented in Section 3.2, and their subtracted counterparts in Section 3.3 and Appendix B. For each type of black hole (as characterized by its mass, rotation and charge parameters) the Green's function at the horizon is independent of the warp factor but the counterterms are not, leading to differing results for the vacuum polarization. In each case, the vacuum polarization is captured by the formula
\begin{equation}\label{gaussian2}
\langle\varphi^2\rangle_{r_+}=\frac{1}{48\pi^2}\left(K_0+R^r_{\,\,r}\right)\,,
\end{equation}
which expresses it in terms of the horizon intrinsic curvature and the Ricci tensor (evaluated at the horizon, or at the horizon pole for rotating black holes).

For static black holes, the results are qualitatively similar in the subtracted and original cases. For rotating black holes, we noted that a sign change which was observed by Frolov \cite{Frolov:1982pi} to occur in the original black holes at high values of $a$ is absent in the subtracted black holes, for which $\langle\varphi^2\rangle$  is always positive. We also confirmed that the subtracted vacuum polarization can be obtained from the original one through a simple scaling limit, and that both results coincide (and vanish) in the BPS limit characterizing static extremal black holes. In section 4 we computed the horizon vacuum polarization for static black hole solutions that interpolate between the original and the subtracted geometry, according to the solution-generating transformations labelled by $\alpha_I$.

Our methods potentially be extended to compute vacuum polarization of the analogues of subtracted geometry that were constructed in \cite{Cvetic:2014sxa} for the Chow-Comp\`ere solution \cite{Chow:2014cca}.
The expressions we provide in Appendix A for the full Green's function on subtracted backgrounds, where it can be computed in closed form, are a promising starting point for further investigations of quantum effects on the subtracted black holes and their comparison with the original ones. Among possible avenues for further research are computations of the stress-energy tensor, numerical investigations of $\langle\varphi^2\rangle$ beyond the horizon, extensions to other fields beyond the minimally coupled massless scalar, and investigations of the self-force problem.

\section{Acknowledgements}
We thank B. Wardell and A.C Ottewill for useful email correspondence.  The work is supported in part by the DOE Grant DOE-EY-76-02- 3071 (MC), the Fay R. and Eugene L. Langberg Endowed Chair (MC), and the Slovenian Research Agency (ARRS) (MC).
% and the Simons Foundation Fellowship (MC). 
  \appendix
\section{Green's Function for subtracted geometry black holes}
  
In this Appendix we describe the computation of the Euclidean Green's function for subtracted black holes. The equation to solve is:
\begin{equation}\label{greeneq}
\Box \; G_H( -i\tau , x , \theta , \phi \; ; {-i \tau}' , {x}' , {\theta}' , {\phi}') = -i \frac{1}{(r_+-r_-)}\frac{1 }{ \sqrt{-g}} \delta ( \tau - {\tau}') \delta ( x-x') \delta ( \Omega , \Omega')\,,
\end{equation}
where $g$ is the determinant of the metric (\ref{metric4d}), and
\begin{equation}\label{xdef}
x = \frac{r - \frac{1}{2}(r_+ + r_-)} { r_+ - r_-}\,.
\end{equation}

 $  \delta ( \Omega , \Omega')$ is the delta function on the two-sphere and can be expanded in terms of the Legendre polynomials as 
\begin{equation}
\delta ( \Omega , \Omega') =\sum_l \frac{( 2 l+1)}{4\pi} P_l ( \cos \Theta)\,,
\end{equation}
where $\Theta$ is the angle between $\Omega$ and $\Omega'$. Likewise the temporal delta function may be expanded:
\begin{equation}
\delta ( \tau - {\tau}') = \frac{\kappa}{ 2 \pi} \sum_{n = -\infty}^ {\infty} \mathrm{e}^{ i n \kappa( \tau -{\tau}')}\,,
\end{equation}
where $\kappa=\kappa_+$, since the Euclidean Green's function must have the periodicity given by the external horizon's surface gravity.

In the static case, where $r_+ - r_-=2m$, the Green's function may be expanded in the following form:
\begin{equation}\label{expansion}
G_H( -i\tau , x , \theta , \phi \; ; {-i \tau}' , {x}' , {\theta}' , {\phi}')= \frac{1}{2m} \frac{i\kappa}{ 2 \pi} \sum_{n =- \infty}^ {\infty} e^{ i n \kappa( \tau -{\tau}')}\sum_l \frac{ (2 l+1)}{4\pi} P_l ( \cos \Theta) G _{ln}(x,x')\,.
\end{equation} 
Substituting this into (\ref{greeneq}) gives us the following equation for the radial Green's function $G_{ln}$:
\begin{equation}
\left[ \frac{\partial}{\partial x} \left(x^2 - 
\frac{1}{4}\right)\frac{\partial}{\partial x}-    
\frac{n^2 }{4\left(x-\half \right)}
  + \frac{1}{4\left( x+\half\right)}
 \left(\frac{n \kappa}{\kappa_-}   \right)^2 -  l(l+1) \right] G_{ln}(x,x')= - \delta( x- x')\, . \label{wave}
\end{equation}
The solution to this equation is constructed from two independent solutions of the corresponding homogeneous  equation:
\begin{equation}
\left[ \frac{\partial}{\partial x} \left(x^2 - 
\frac{1}{4}\right)\frac{\partial}{\partial x}-    
\frac{n^2 }{4\left(x-\half \right)}
  + \frac{1}{4\left( x+\half\right)}
 \left(\frac{n \kappa}{\kappa_-}   \right)^2 -  l(l+1) \right]  \chi_{ln}(x)= 0\, . \label{wavehom}
\end{equation}
The solutions to this equation have been derived in \cite{Cvetic:2013lfa,Cvetic:2014ina} and are expressed in terms of hypergeometric functions $F(a,b,c;z)$. One has two independent solutions $ \chi_{ln}^{(1)}$,  $ \chi_{ln}^{(2)}$, of which the  first one is regular at the horizon ($x=\frac{1}{2}$) and the second one is regular  at infinity $(x\to+\infty)$. These  solutions are
\begin{eqnarray}
\chi_{ln}^{(1)} &=& \left(x+\half\right)^{- (l+1) }
\left( \frac{x-\half}{x+\half } \right) ^{  \frac{n}{2 }}F \left(a_{ln},b_{ln}, 1 +n
; \frac{x-\half}{x+\half } \right) \,, \nonumber\\
\chi_{ln}^{(2)} &=& \left(x+\half\right)^{- (l+1) }
\left( \frac{x-\half}{x+\half } \right) ^{  \frac{n}{2 }}
F \left(a_{ln},b_{ln}, ,2 l + 2
; \frac{1}{x+\half } \right) \,,
\end{eqnarray}
with
\begin{eqnarray}
a_{ln}&=&l+1+\frac{n}{2}\left(1+\frac{\kappa}{\kappa_-}\right)\,,\nonumber\\
b_{ln}&=&l+1+\frac{n}{2}\left(1-\frac{\kappa}{\kappa_-}\right)\,.
\end{eqnarray}

The formula for Green's function is simply

\begin{equation}
G_{ln} ( x,x') =  \frac{\Gamma\left(a_{ln}\right) \Gamma\left(b_{ln}\right) }{  (2l+1)!\, n!} \,\left[\chi_{ln}^{(1)}( x') \,\chi_{ln}^{(2)}( x)\,H(x-x') + \chi_{ln}^{(1)}( x) \,\chi_{ln}^{(2)}( x') H(x'-x) \right]\,,
\end{equation}
where $H(x)$ is the Heaviside step function. The prefactor is obtained from the Wronskian of the two solutions:
\begin{equation}
\left(x'^2-\frac{1}{4}\right)  W_{(\chi_{ln}^{(1)} , \chi_{lm}^{(2)})}(x') = - \frac{  (2l+1)!\, n!}{\Gamma\left(a_{ln}\right) \Gamma\left(b_{ln}\right) }\,.
\end{equation}

The solutions for $n=0$  are written more simply in terms of Legendre functions:
\begin{eqnarray}
\chi^{(1)}_{l0}(x) & = & P_l(2x)\,,\nonumber\\
\chi^{(2)}_{l0}(x) & = & \frac{(2l+1)!}{2(l!)^2}Q_l(2x)\,.
\end{eqnarray}
Thus the radial Green function for $n=0$ is given simply by
\begin{equation}\label{green0}
G_{l0}(x,x')=G_{0l0}(x,x')=2\left[P_l(2x)Q_l(2x')H(x'-x)+P_l(2x')Q_l(2x)H(x-x')\right]\,,
\end{equation}
as we claimed in Section 3.

In the rotational case  the ansatz expression for the Green's function is\footnote{In this expression and throughout the rest of Appendix A, $m$ will always stand for the index labeling $\phi$-dependent modes and not for the black hole mass parameter.}
\begin{align}\label{greenexpandrot}
G_H( -i\tau , x , \theta , \phi \; ; {-i \tau}' , {x}' , {\theta}' , {\phi}')&= \frac{1}{r_0} \frac{ i \kappa}{ 2 \pi} \sum_{n =- \infty}^ {\infty} e^{ i n \kappa( \tau -{\tau}')} \sum_{l=0}^{\infty}\frac{(2l+1)}{4\pi} \nonumber\\
&\times\sum_{m =- l}^ {l} \frac{(l-m)!}{(l+m)!}e^{ i m ( \phi -{\phi}')} P_l^m( \cos{\theta})P_l^m( \cos{\theta'}) G _{mln}(x,x')\,,
\end{align} 
where $r_0=r_+-r_-$.  Upon substitution in (\ref{greeneq}), we use the delta function expansion
\begin{equation}
\delta(\Omega,\Omega')= \sum_{l=0}^{\infty}\frac{(2l+1)}{4\pi} \sum_{m =- l}^ {l} \frac{(l-m)!}{(l+m)!}e^{ i m ( \phi -{\phi}')} P_l^m( \cos{\theta})P_l^m( \cos{\theta'})\,, 
\end{equation}
and obtain the following radial equation to solve
\begin{align}
\,&\Bigg[ \frac{\partial}{\partial x} \left(x^2 - 
\frac{1}{4}\right)
\frac{\partial}{\partial x}-    
\frac{1}{4\left(x-\half \right)}\left( n -  \frac{m\Omega_+}{\kappa}  \right)^2 
  + \frac{1}{4\left( x+\half\right)}
 \left(\frac{n \kappa- m\Omega_-}{\kappa_-}   \right)^2 \nonumber\\
 &-  l(l+1) \Bigg] G_{mln}(x,x')= - \delta( x- x')\, , \label{wave}
\end{align}
Now we proceed as before and solve the homogeneous equation first. The equation is of the same essential form as the static one (\ref{wavehom}), so the solutions will take the same form. We must be careful, however, to write the solutions in terms of the absolute values of the combinations of parameters that appear squared in the equation. The solutions are written as:
\begin{eqnarray}
\chi_{mln}^{(1)}(x) &=& \left(x+\half\right)^{- (l+1) }
\left( \frac{x-\half}{x+\half } \right ) ^{ \frac{1}{2}\left|n  - \frac{m \Omega _+}{\kappa}\right|}
F \left( a_{mln},
b_{mln}, 1 +  \left| n  - \frac{m \Omega_+}{\kappa}\right|; \frac{x-\half}{x+\half } \right) \,,\nonumber\\  
\chi_{mln}^{(2)}(x) &=& \left(x+\half\right)^{- (l+1) }
\left( \frac{x-\half}{x+\half } \right ) ^{ \frac{1}{2}\left|n  - \frac{m \Omega _+}{\kappa}\right|}
F \left( a_{mln},
b_{mln}, 2l+2; \frac{1}{x+\half } \right) \,,  
\end{eqnarray}
where 
\begin{eqnarray}
a_{mln}&=& l + 1 + \frac{ 1}{ 2 } \left| n  - \frac{m \Omega_+}{\kappa}\right| + \frac{ 1}{ 2 } \left| \frac{n \kappa  - m \Omega_-}{\kappa_-}\right|\,, \nonumber\\
b_{mln}&=& l + 1 + \frac{ 1}{ 2 } \left| n  - \frac{m \Omega_+}{\kappa}\right| - \frac{ 1}{ 2 } \left| \frac{n \kappa  - m \Omega_-}{\kappa_-}\right|\,.
\end{eqnarray}

The full radial Green's function is then:
\begin{align}
G_{mln} ( x,x') &= \frac{ \Gamma( a_{mln}) \Gamma(b_{mln})} {(2l+1)!\Gamma\left(1 +  \left| n  - \frac{m \Omega_+}{\kappa}\right| \right) }\nonumber\\
&\times \left[ \chi_{mln}^{(1)}( x') \chi_{mln}^{(2)}( x) H(x'-x)+\chi_{mln}^{(1)}( x) \chi_{mln}^{(2)}( x') H(x-x')\right]\,.
\end{align}

It is easily verified that in the static case $G_{ln}(x,x')=0$ when $x=1/2$ unless $n=0$. In the rotating case, $G_{mln}(x,x')=0$  when $x=1/2$ unless $n\kappa = m\Omega_+$. Setting aside the possibility of the black hole parameters being fine-tuned to make $\kappa/\Omega_+$ a rational number, this will only happen when $n=m=0$. This justifies our claim that only the  $G_{l0}(x,x')$ and $G_{0l0}(x,x')$ are relevant for the horizon vacuum polarization.

\section{Vacuum polarization for the original four-charge rotating black hole}

In this Appendix we present the general result for the vacuum polarization at the horizon pole $(r=r_+,\theta=0)$ of the non-subtracted rotating black hole with four distinct charges. We use the following notation to make the result more compact:
\begin{eqnarray}
d_1 &=& 2\sum_{I}\sinh^2\delta_I\,,\\
d_2 &=& 4 \sum_{I< J}\sinh^2\delta_I \sinh^2\delta_J\,,\\
d_3 &=&  8\sum_{I<J<K}\sinh^2\delta_I \sinh^2\delta_J\sinh^2\delta_K\,,\\
d_4 &=& 8\,\Pi_c \Pi_s-4\prod_{I<J<K}\sinh\delta_I\sinh\delta_J\sinh\delta_K\,,\\
d_5 &=& (4\Pi_s)^2\,,
\end{eqnarray}
and in addition we write as before $r_0=2\sqrt{m^2-a^2}$.
The full result can be written as:
\begin{equation}
\langle\varphi^2\rangle_{r_+,\,\theta=0}^{gen_{orig}}=\frac{A}{48\pi^2m^2r_0\,C^{3/2}}+\frac{B}{192\pi^2\,C^{5/2}}\,,
\end{equation}
where:
\begin{equation}
A=4(4+d_1)\left(m^2-a^2\right)^2-m\,r_0\left(a^2(12+4d_1+d_2+d_4)+m^2(-8-2d_1+d_3+d_5)\right)\,,
\end{equation}
\begin{align}
B&=-2m\,a^2\left[12d_1^2+6d_2^2+d_3(16+d_3-2d_4)-2d_5(d_4-4)+2d_2(12+3d_3-d_4+d_5)\right.\nonumber\\
&\left.+2d_1(12+9d_2+5d_3-d_4+2d_5)\right]-a^2r_0\left(2d_1+2d_2+d_3\right)\left(4+2d_1+d_2-d_4\right)\nonumber\\
&+2m^3\left[16d_1^2+8d_2^2+3d_3(8+d_3)+4d_5(4+d_3)+2d_5^2+2d_2(16+5d_3+3d_5)\right.\nonumber\\
&\left.+2d_1(16+12d_2+8d_3+5d_5)\right]+m^2r_0\left[12d_1^2+8d_2^2+16d_5+3d_3(8+d_3+d_5) \right.\nonumber\\
&\left.+2d_2(16+5d_3+3d_5)+2d_1(16+12d_2+8d_3+5d_5)\right]\,,
\end{align}
\begin{equation}
C=m^2(8+4d_1+2d_2+d_3+d_5)-a^2(4+2d_1+d_2-d_4)+m\,r_0(4+2d_1+d_2+\frac{1}{2}d_3)\,.
\end{equation}

\end{document}